\documentclass[conference,10pt]{IEEEtran}
\IEEEoverridecommandlockouts

\usepackage{url}
\usepackage[utf8]{inputenc}
\usepackage{amsmath,amssymb}

\usepackage{amsmath,amssymb,amsthm}
\usepackage{enumerate}
\usepackage{stfloats}
\usepackage{comment}
\usepackage{bbm}
\usepackage{siunitx}
\usepackage{mathtools}
\usepackage{accents,latexsym,cancel}
\usepackage{cite}

\usepackage[dvipsnames]{xcolor}
\definecolor{myPink}{RGB}{255,105,183}

\usepackage[T1]{fontenc}
\usepackage{graphics} 
\usepackage{epsfig} 
\usepackage[mathscr]{euscript}
\usepackage{algorithm}
\usepackage[noend]{algpseudocode}
\usepackage{bbm}
\makeatletter
\def\BState{\State\hskip-\ALG@thistlm}
\makeatother

\usepackage{tikz}
\usetikzlibrary{arrows,shapes,chains,matrix,positioning,scopes,patterns,calc,
decorations.markings,
decorations.pathmorphing,
}

\usepackage{pgfplots}
\pgfplotsset{compat=1.3}
\usepgflibrary{shapes}

\renewcommand{\epsilon}{\varepsilon}

\newcommand{\RNum}[1]{\uppercase\expandafter{\romannumeral #1\relax}}

\newcommand{\av}{\ensuremath{\mathbf{a}}}

\newcommand{\cv}{\ensuremath{\mathbf{c}}}

\newcommand{\dv}{\ensuremath{\mathbf{d}}}
\newcommand{\gv}{\ensuremath{\mathbf{g}}}
\newcommand{\hv}{\ensuremath{\mathbf{h}}}

\newcommand{\mv}{\ensuremath{\mathbf{m}}}
\newcommand{\nv}{\ensuremath{\mathbf{n}}}
\newcommand{\pv}{\ensuremath{\mathbf{p}}}
\newcommand{\qv}{\ensuremath{\mathbf{q}}}
\newcommand{\rv}{\ensuremath{\mathbf{r}}}
\newcommand{\sv}{\ensuremath{\mathbf{s}}}
\newcommand{\Sv}{\ensuremath{\mathbf{S}}}

\newcommand{\xv}{\ensuremath{\mathbf{x}}}
\newcommand{\yv}{\ensuremath{\mathbf{y}}}
\newcommand{\zv}{\ensuremath{\mathbf{z}}}

\newcommand{\Am}{\ensuremath{\mathbf{A}}}
\newcommand{\Dm}{\ensuremath{\mathbf{D}}}

\DeclareMathAlphabet{\mcl}{OMS}{cmsy}{m}{n}

\DeclareMathOperator*{\argmin}{\,arg\ min}

\newlength\tikzwidth
\newlength\tikzheight

\definecolor{mycolor1}{rgb}{0.63529,0.07843,0.18431}%
\definecolor{mycolor2}{rgb}{0.00000,0.44706,0.74118}%
\definecolor{mycolor3}{rgb}{0.00000,0.49804,0.00000}%
\definecolor{mycolor4}{rgb}{0.87059,0.49020,0.00000}%
\definecolor{mycolor5}{rgb}{0.00000,0.44700,0.74100}%
\definecolor{mycolor6}{rgb}{0.74902,0.00000,0.74902}%

{\begin{list}{$\bullet$}
   {\setlength{\itemsep}{0in}
    \setlength{\labelsep}{0.05in}
    \setlength{\labelwidth}{0.2in}
    \setlength{\leftmargin}{0.25in}
    \setlength{\rightmargin}{0in}
    \setlength{\topsep}{0in}
   }}
{\end{list}}
\newcommand{\T}{^{\mbox{\tiny T}}}

\newcommand{\Pm}{\mathbf{P}}

\newcommand{\Ym}{\mathbf{Y}}

\newcommand{\Bm}{\mathbf{B}}
\newcommand{\Zm}{\mathbf{Z}}
\newcommand{\Sm}{\mathbf{S}}

\newcommand{\Gm}{\mathbf{G}}
\newcommand{\pmd}{\text{P}_{\mathrm{md}}}
\newcommand{\pfa}{\text{P}_{\mathrm{fa}}}
\newcommand{\pe}{\ensuremath{\text{P}_{\mathrm{e}}}}

\title{Scalable Cell-Free Massive MIMO  Unsourced Random Access}
\author{Michail Gkagkos, Jean-Francois Chamberland, Costas N. Georghiades, Krishna R. Narayanan \\
Department of Electrical and Computer Engineering, Texas A\&M University\\
Email: \{gkagkos, chmbrlnd, georghiades, krn\}@tamu.edu
\thanks{
This material is based upon work supported, in part, by the National Science Foundation (NSF) under Grants CCF-2131106 \& CNS-2148354, and by Qualcomm Technologies, Inc., through their University Relations Program.}
}
\begin{document}

\maketitle

\begin{abstract}
Cell-Free Massive MIMO systems aim to expand the coverage area of wireless networks by replacing a single high-performance Access Point (AP) with multiple small, distributed APs connected to a Central Processing Unit (CPU) through a fronthaul.
Another novel wireless approach, known as the unsourced random access (URA) paradigm, enables a large number of devices to communicate concurrently on the uplink.
We consider a quasi-static Rayleigh fading channel paired to a scalable cell-free system, wherein a small number of receive antennas in the distributed APs serve devices equipped with a single antenna each.
The goal of the study is to extend previous URA results to more realistic channels by examining the performance of a scalable cell-free system.
To achieve this goal, we construct a coding scheme that adapts the URA paradigm to various cell-free scenarios. 
Empirical evidence suggests that using a cell-free architecture can improve the performance of a URA system, especially when taking into account large-scale attenuation and fading.
\end{abstract}

\section{Introduction}
Massive Machine Type Communication (mMTC) and Cell-Free (CF) systems have been proposed for future wireless communication systems~\cite{5Gsurvey,6Gsurvey}.
The mMTC paradigm seeks to enable connectivity at scale, whereas CF systems aim to expand coverage. 
To achieve these goals, the traditional base station located in the middle of the coverage area is substituted by many small access points (APs), each with a limited number of antennas.
These APs are distributed throughout the geographical region of interest.
They are linked to a common CPU, which is tasked with signal aggregation and network coordination.
As a result, the probability that a mobile device finds itself at the cell edge is greatly reduced.
Defining characteristics of emerging mMTC devices, such as sensors, include lower transmit power and short battery lives.
Thus, these technologies appear naturally suited for this type of traffic.
The objective of this study is to develop a scalable~\cite{Scalable-Cell-Free} cell-free system that can support machine-type communications by leveraging the unsourced random access (URA) paradigm~\cite{yuryAWGN}.

\subsection{Unsourced Random Access Channels}
URA offers a different approach to random access for handling the communication needs of devices that operate without direct supervision. 
This model has become a popular framework for IoT wireless networks. 
The motivation behind uncoordinated access is that, as the number of potential users increases, it becomes difficult to allocate resources based on the length of queues and channel conditions~\cite{3GPP-R16-MAC,uplinkScheduling}.
This situation becomes especially challenging when devices send short packets only sporadically. 
A practical solution is to have all active devices share a same codebook, so that the system can operate irrespective of the total number of devices and focusing instead exclusively on the active population.
In such scenarios, the system objective is to recover the set of sent messages, regardless of which devices sent them.
If a device wishes to reveal its identity, it can embed it in the payload of its own message.
Researchers in this area have been designing coding schemes for additive white Gaussian noise (AWGN) channels \cite{avinash2017,SKP,AsitPolar,VamsiCCSAWGN}, quasi-static SISO fading channels \cite{AndreevPolarEM,YuryfinitePayload,AlexeyRecoverableCodes,YuryEnergyEfficientMAC},  quasi-static MIMO \cite{LiuTanner,UstinovaMIMO}, and Massive MIMO fading channels\cite{fengler2022pilotbased,gkagkosSpawc,TolgaOrthogonal2022, MaximeTensor, tolgaFASURA}.
Furthermore, Shao et al.\ explore realistic channels and practical considerations~\cite{firstCellFree-URA}.
The authors therein study a coordinated CF system and they design an algorithm, called cooperative
activity detection (CAD), to identify active devices.
Furthermore, they show that CAD can be applied to the URA setting as well.
They borrowed  ideas from Coded Compressed Sensing (CCS) in \cite{VamsiCCSAWGN}, and adopt their algorithm as part of the decoder for the ensuing CCS scheme.

\subsection{Cell-Free Massive MIMO}
\label{subsec:CF}

\begin{figure}
    \centering
\includegraphics[scale=0.15]{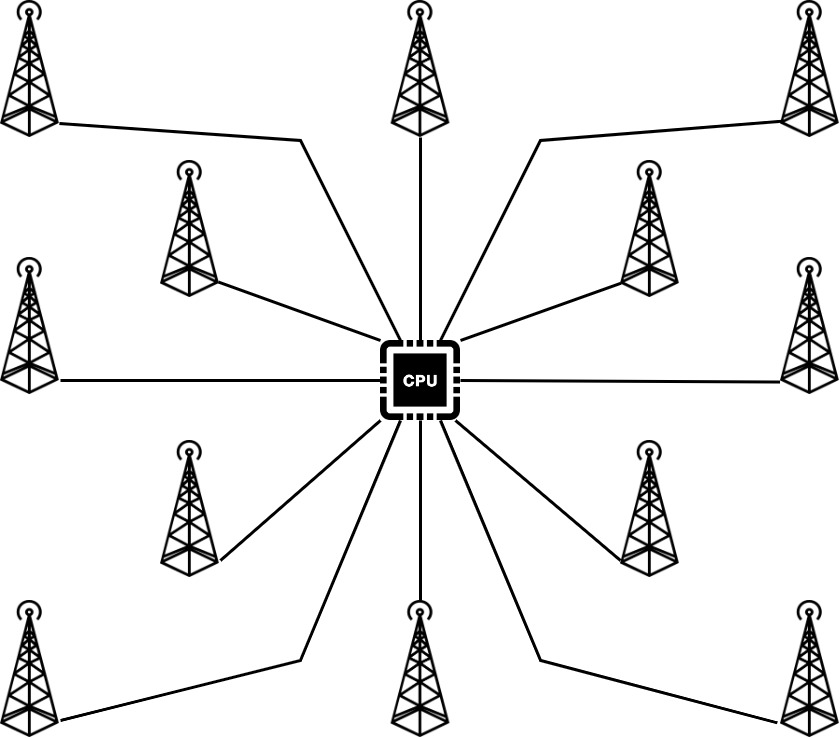}
    \caption{Example of a CF system with 12 APs and one CPU.}
    \label{fig:exampleCF}
\end{figure}

A new system architecture, called cell-free massive MIMO, has been proposed for next-generation wireless communication systems \cite{VersusSmallCells, Cell-FreeFirstPaper}.
The idea is to remove the basestation located in the middle of a cell and distribute antennas within the same cell geographic area.
These distributed APs are connected to the CPU via a fronthaul.
This distributed architecture, with multiple rudimentary APs, can together serve a larger number of user equipment (UE).
Each AP features $N$ antennas (a small number).
Moreover, for the purpose of exposition, each UE has a single antenna~\cite{bookUC-CF}. 
Figure~\ref{fig:exampleCF} contains a notional diagram for a CF system.

In such a distributed wireless network, there are four different levels of cooperation between APs and CPU \cite{MakingCompetitive}.
At \textit{Level~4}, all APs send the received signals to the CPU, which then performs channel estimation and symbol detection.
On the other hand, at \textit{Level~3}, channel estimation and data detection are performed at the APs; the CPU gathers the local estimates from all APs and it makes the final decisions using a linear detector that is solely dependent on the channel statistics.
The cooperation mechanism at \textit{Level~2} is a simplified version of \textit{Level~3} where the CPU computes the sum of the local estimates.
The most distributed setting occurs at \textit{Level~1} where detection is performed independently at every AP, which in turn serves at most one UE.
For the latter scenario, there is no information exchange between APs and the CPU.

\noindent
\textbf{Scalable  Cell-Free System:}
The initial proposals for CF systems suggests that all UEs be served by all APs \cite{Cell-FreeFirstPaper, Cell-Free-mMIMOsys}.
However, this approach is impractical and unnecessary when the network covers a large geographic area.
Instead, each UE is in close physical proximity to only a small number of APs; and an AP serves a UE only if its signal power is significant compared to thermal noise.
This latter approach lowers computational complexity and reduces fronthaul links to the CPU, thereby making the system scalable \cite{Scalable-Cell-Free}.
Existing literature on CF systems discusses dynamic cooperation clustering (DCC) to capture the association between APs and UEs.
We do not discuss this aspect in the present article because it is only applicable in the context of coordinated access~\cite{DCCfirstPaper}\footnote{In cell-free unsourced random access channels, the APs are blind to the identities of the users.}.
The notion of \textit{scalability} is defined in \cite{Scalable-Cell-Free,bookUC-CF} as follows:
when the coverage area of a network expands, it is crucial to ensure that the technology is scalable, which means that additional UEs and/or APs can be integrated into the network without requiring the existing infrastructure to be upgraded.
For example, if the geographic area remains constant but the number of UEs increases, more APs can be deployed to serve the additional UEs without affecting the computational capacity or maximum fronthaul capacity of the existing APs. 

\subsection{Main Contributions}
As mentioned above, the literature on CF-URA is sparse, with one candidate scheme found in \cite{firstCellFree-URA}. 
Several key aspects of such systems are yet to be studied and the system model developed in this article differs significantly from established results.
For instance, in \cite{firstCellFree-URA}, each AP is linked to several nearby APs via fronthaul links, and two APs can communicate only if they are one-hop neighbors, thereby reducing the communication load; effectively, there is no CPU in the system. 
In contrast, in our article, the APs do not share data with each other; rather the CPU collects data estimates from all APs. 
Furthermore, Shao et al.\ \cite{firstCellFree-URA} consider APs with a large number of antennas (100--300).
We consider distributed APs with a very small number of antennas.

More specifically, we construct a scheme called CEFURA (Cell-Free Unsource Random Access), which offers a \textit{Level~2} implementation for a scalable CF system.
Herein, each AP recovers a subset of the active UEs; the AP estimates each channel and detects the data of nearby UEs.
Local estimates are subsequently transferred to the CPU for final decisions.
We evaluate the performance of CEFURA and compare it to a centralized version of the same system, where only one high-performance access point serves the same area.
Our goal is to demonstrate the benefits of a CF system, and motivate other researchers to consider more practical URA channels.
Additionally, we examine the performance of our scheme for various UE location distributions through simulations.

\noindent
\textbf{Notation:}
Throughout, $\mathbb{C}$ refer to complex numbers, and we use $[n]$ to denote $\{1,2,\dots,n\}$.
We employ boldface lowercase $\av$ and boldface uppercase letters $\Am$ to indicate vectors and matrices.
The matrices $\Am\T$ and $\Am^*$ represent the transpose and the conjugate transpose of matrix $\Am$.
Sets are labeled with calligraphic letters, e.g., $\mathcal{A}$.
We also adopt a programming-style notation with $\Am[:,t]$ and $\Am[k,:]$ representing the $t$th column and $k$th row of $\Am$, respectively.
We use $\| \cdot \|_\mathrm{F}$ and $\| \cdot \|_2$ for the Frobenius and second norms, respectively.

\section{System Model}
We consider an uplink cell-free system with $K_{\mathrm{tot}}$ UEs, and $M$ APs randomly located in area of $D \times D  \ \mathrm{m}^2$ .
UEs have a single antenna and the APs are equipped with an Uniform Linear Array (ULA) with  $N$ antenna elements. 
We assume that the antenna spacing is 0.5m, and the array response  is given by,
\begin{align*}
    \qv(\phi) =\begin{bmatrix}
        1 & e^{j\pi\sin(\phi)} & e^{j2\pi\sin(\phi)} & \dots & & e^{j(M-1)\pi\sin(\phi)}
    \end{bmatrix},
\end{align*}
where $\phi$ is the angle of arrival.
We assume that there is a single CPU in the network to which all $M$ APs are connected through a fronthaul; wired links are taken to have infinite capacity.
In each time slot, $K \ll K_{\mathrm{tot}}$ devices are active, each aiming to transmit their data to the APs.
We also assume perfect frame synchronization.

\subsection{Channel Model}
\label{subs:channel}
The channel between the $m$th AP and the $k$th UE has two components, the large and the small scale coefficients; it is defined by
\begin{align*}
    \gv_{k,m} = \beta_{k,m}^{\frac{1}{2}}\hv_{k,m} .
\end{align*}
The large scale coefficient $\beta_{k,m}$ is a function of the path loss and the shadow fading.
The elements of the small scale component $\hv_{k,m}$, i.e. $h_{k,m,n}$, are generated assuming ULA  (see \cite{emilBookMassiveMIMO}).
We assume a quasi-static Rayleigh fading model whereby channel coefficients remain fixed during 
the entire transmission.
For the path loss, we use the following 3GPP urban microcell propagation model in \cite[Table B.1.2.1-1]{3GPP-R9}, (also used in \cite{MakingCompetitive}),  with a carrier frequency of 2~GHz,
\begin{align}
    \beta_{k,m}[\mathrm{dB}]  = -30.5 -35\log_{10}d_{mk} + F_{km}, \label{eq:LSC}
\end{align}
where $d_{km}$ is the distance between the $m$th AP and the $k$th UE, and $F_{km} \sim N(0,16)$ is the shadow fading.
It should be noted that the shadow fading is correlated from an AP to different UEs as \cite[Table B.1.2.2.1-4]{3GPP-R9} and their correlation is given by
\begin{align*}
    \mathbb{E}[F_{km}F_{ij}] =
    \begin{cases}
    16\times 2^\frac{-\Delta X_{ki}}{9}, \ \ \text{if $ l = j$} \\
    0, \ \ \text{otherwise}.
    \end{cases}
\end{align*}
Above, $\Delta X_{ki}$ denotes the distance between UE~$k$ and UE~$i$.
We note that the correlation of shadowing effects between two adjacent APs is negligible within our simulations.

\subsection{Received Signal}
Let $\mv_k$ be the $B$-bit message of UE~$k$, and $\xv_k = \mathcal{E}(\mv_k) \in \mathbb{C}^n$ be the encoded and modulated signal (input to the channel) corresponds to message $\mv_k$.
Then, the received signal at the $N$ receive antennas of the $m$th AP takes the form
\begin{equation} \label{eq:systemSum}
\begin{split}
    \Ym_m &= 
    \sum_{k \in {\cal K}} \xv(\mv_k) \gv_{k,m}\T + \Zm_m ,
\end{split}
\end{equation}
where $\Ym_m \in \mathbb{C}^{n \times N}$ and the set of the active UEs is labeled $\mathcal{K}$.
The vector $\gv_{k,m} \in \mathbb{C}^{N}$ is the combination of the large and small channel coefficients, as described in Section~\ref{subs:channel}, from the $k$th user to the $m$th access point.
Additive noise component $\Zm_m \in \mathbb{C}^{n \times N}$ is a matrix with i.i.d.\ entries, each drawn from a circularly symmetric complex Gaussian distribution ${\cal CN}(0,\sigma^2)$.
Furthermore, every transmit signal must satisfy power constraint $(1/n)\| \xv(\mv_k) \|^2 \leq P_k$.
At the CPU, the decoder aims to produce a set $\hat{{\cal K}}$ of candidate messages with cardinality at most $K$.
The system performance is evaluated in terms of probability of missed detection $\pmd$ and probability of false alarm $\pfa$.
For the problem at hand, these two error probabilities are given by
\begin{xalignat*}{2}
\pmd &= \frac{\mathbb{E}[n_{\mathrm{ms}}]}{K} & \pfa &= \mathbb{E}\left[ \frac{n_{\mathrm{fa}}}{\hat{K}} \right]
\end{xalignat*}
where $\hat{K}$ is the number of recovered users, and $n_{\mathrm{ms}}$ and $n_{\mathrm{fa}}$ denote the number of misses and false alarms, respectively.
We define the \textit{Error Rate} as $\pe =  \pmd + \pfa$.

\section{CEFURA}
In this section we describe the main components of CEFURA, beginning with the UE design, then the AP structure, and finally with the channel decoder at the CPU.

\subsection{User Equipment (UE)}
Each device splits its $B$-bit message $\mv$ into two parts, i.e., $\mv = [ \mv_f \ \mv_s]$, with lengths of $B_f$ and $B_s$, respectively.

\subsubsection{Encoding $\mv_f$}
Let $\Pm$ and $\Am$ represent the master sets of pilots and spreading sequences. 
The coefficients of matrix $\Am$ are distributed as complex Gaussian random variables, i.e., $a_{j} \sim {\cal CN}(0,1)$.
The columns of matrix $\Am$ can be viewed as spreading sequences of length $L$, and they are normalized to have energy of $L$.
There are $J = 2^{B_f}$ possible spreading sequences attached to every time instant.
Likewise, let ${\bf P} \in \big\{ \pm \frac{1}{\sqrt{2}} \pm \frac{j}{\sqrt{2}} \big\}^{n_p \times J}$ be a matrix whose columns are possible pilot sequences.
The process of choosing which spreading sequences and pilots to use involves function $\phi : \ \{0,1\}^{B_f} \rightarrow [J]$, which takes the binary message $\mv_f$ as input and maps it to an index in the range $[J]$. 
Thus, if the initial part of the message to be transmitted is $\mv_f$, the user will utilize spreading sequence $\Am[:,\phi(\mv_f)]$ and pilot sequence $\Pm[:,\phi(\mv_f)]$ corresponding to the index obtained through function $\phi$.
The overall encoding function for $\mv_f$ can be summarized as 
\begin{equation*}
    g(\mv_f) \rightarrow \left( \Am[:,\phi(\mv_f)],  \Pm[:,\phi(\mv_f)] \right).
\end{equation*}
There is no guarantee that active users will each pick a unique sequence from an orthogonal subset. 
Instead, active users pick sequences randomly from a collection of possibly non-orthogonal sequences. 
Under the URA framework, it is not feasible to choose sequences manually, as two devices with the same message will unavoidably transmit identical signals. 
Nevertheless, it is possible to reduce the chances of collisions by increasing the length of binary message $\mv_f$.

\subsubsection{Encoding $\mv_s$}

The second part of the message, namely $\mv_s$, is first encoded using a cyclic redundancy check (CRC) code.
The resulting codeword of length $B_{c} = B_s + B_{\mathrm{crc}}$  acts as input to an encoder for a $(n_c,B_c)$ polar code with $n_c-B_c$ frozen bit positions.
Suppose $\cv \in \{0,1\}^{n_c}$ is the output of the polar encoder, then $\cv$ is modulated using QPSK to obtain vector $\sv$ of length $T = n_c/2$.
Finally, the QPSK symbols, $\sv$, are spread using the $\phi(\mv_f)$th column of $\Am$.
The resulting signal $\dv$ can be expressed as 
\begin{equation} \label{eq:secondPart}
    \dv(\mv_f,\mv_s) = \sv 	\otimes \av_{\phi(\mv_f)}
\end{equation}
where $	\otimes$ is the Kronecker product, and $\av_{\phi(\mv_f)} = \Am[:,\phi(\mv_f)]$.
The input signal to the channel is the concatenation of the pilot sequence $\pv(\mv_{f})$ and spread codeword $\dv(\mv_f,\mv_s)$. 
Altogether, when the message of user~$k$ is $\mv_k = \left( \mv_{k,f}, \mv_{k,s} \right)$, 
the signal sent by this user is equal to
\begin{equation*}
\xv_k = \sqrt{P_k} \begin{bmatrix} 
\pv\T(\mv_{k,f}) &
\dv\T(\mv_{k,f},\mv_{k,s}) \end{bmatrix}\T
\end{equation*}
where $\sqrt{P_k}$ is the transmit power, $\pv(\mv_{k,f}) = \Pm[:, \phi(\mv_{k,f})]$ and note  that $\Vert \xv_k \Vert^2 = nP_k$.
With this procedure, the system model of \eqref{eq:systemSum} can be written as the concatenation of
\begin{xalignat}{3}
\Ym^p_m &=  \Pm_a \boldsymbol\Pi_k^{\frac{1}{2}} \Gm_m + \Zm^p_m &\text{and}&&
\Ym^d_m &= \Dm_a \boldsymbol\Pi_k ^{\frac{1}{2}}\Gm_m + \Zm^d_m \nonumber
\end{xalignat}
or, in vector form,
\begin{equation*}
\begin{bmatrix} \Ym^p_m \\ \Ym^d _m\end{bmatrix}
= \begin{bmatrix}  \Pm_a \\
\Dm_a \end{bmatrix} \boldsymbol\Pi_k^{\frac{1}{2}} \Gm_m
+ \begin{bmatrix} \Zm^p_m \\ \Zm_m^d \end{bmatrix}, \ \ \forall m\in [M]
\end{equation*}
where $\Ym^p_m \in \mathbb{C}^{n_p \times N}$, $\Ym_m^d \in \mathbb{C}^{TL \times N}$, and subscript $a$ indicates sub-matrices with active columns only.
That is, the $k$th column of $\Pm_a$ is $\Pm[:,\phi(\mv_{k,f})]$ and the $k$th column of $\Dm_a$ is $\dv_k$, and $\boldsymbol\Pi_k$ is a diagonal matrix with elements $P_k$.
Since there is no coordination between APs and UEs, we assume that all users transmit with the same power, i.e., $P_k = P, \ \forall k \in \mathcal{K}$.
Finally, the $k$th column of $\Gm_m$ corresponds to the channel  between the $k$th UE and the $N$ received antennas at the AP.



\begin{figure}
    \centering
    \scalebox{0.7}{\definecolor{myColor}{RGB}{36, 173, 71}
\begin{tikzpicture}
  [
  font=\normalsize, draw=black, line width=1pt,
  block/.style={rectangle, minimum height=10mm, minimum width=15mm,
  draw=black, fill=gray!10, rounded corners},
  sum/.style={rectangle, minimum height=15mm, minimum width=15mm, draw=black, rounded corners, fill=gray!10},
  message/.style={rectangle, minimum height=5.5mm, minimum width=80mm, draw=black, rounded corners},
  submessage/.style={rectangle, minimum height=5.5mm, minimum width=40mm, draw=black, rounded corners},
  inputCh/.style={rectangle, minimum height=5.5mm, minimum width=20mm, draw=black, rounded corners},
  dataCh/.style={rectangle, minimum height=5.5mm, minimum width=45mm, draw=black, rounded corners}
  ]

\node[message,align=center] (msg)  {Message $\mv$};

\node[submessage,left of=msg, left of=msg,below of=msg, node distance = 1cm] (mf) {$\mv_f$};
\node[submessage,right of= mf, node distance = 4.2cm] (ms) {$\mv_s$};
\draw[->] (-2,-0.25) -- (mf);
\draw[->] (2.2,-0.25) -- (ms);

\node[block,below of= mf, node distance = 1cm] (phi) {$\phi(\cdot)$};
\draw[->] (mf.south) -- (phi.north);

\node[block,left of= phi, left of= phi, below of= phi, node distance = 1.5cm] (P) {$\mathbf{P}$};

\node[block,  right of= P, node distance = 4.5cm] (AT) {$\mathbf{A}$};
\draw[->] (phi.west) -| (P.north)node[midway,above]{$\phi(\mv_f)$};

\draw[->] (phi) -| (AT.north)node[midway,above]{$\phi(\mv_f)$};

\node[block, below of= ms, node distance = 1.8cm] (encoder) {Channel Encoder};
\draw[->] (ms.south) -- (encoder.north);

\node[block, below of= encoder, node distance = 2cm] (qpsk) {QPSK};
\draw[->] (encoder.south) -- (qpsk.north)node[midway,right]{$\cv$};

\node[inputCh, below of= P, node distance = 1.3cm] (pilots) {$\mathbf{P}[:,\phi(\mv_f)]$};
\draw[->] (P.south) -- (pilots.north);

\node[dataCh, left of= qpsk, node distance = 3.9cm] (kron) {$\sv 	\otimes \av_{\phi(\mv_f)}$};

\draw[->](qpsk.west) -- (kron.east) node[midway,above]{$\sv$};

\draw[->](AT.west) -| (kron.north) node[midway,left]{$\av_{\phi(\mv_f)}$};

\end{tikzpicture}}
     \caption{This block diagram highlights the main functionalities of User Equipment transmitter.}
    \label{fig:UEs}
\end{figure}
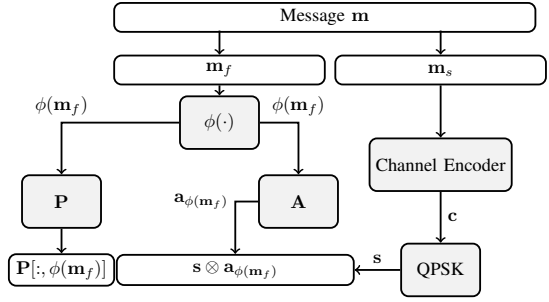
\subsection{Access Point (AP)}

To make the system scalable~\cite{Scalable-Cell-Free}, we have each AP process at most $R_m < K$ UEs.
We consider cooperation at \textit{Level~2} between APs and the CPU.
As a consequence, APs estimate the channel and detect the symbols of each UE, and then send the estimates of the symbols to the CPU.

\subsubsection{Pilot Detector -- Channel Estimation Algorithm}
Since the pilots are not known a priori to APs, pilot detection is necessary to perform channel estimation.
It should be mentioned that since the pilots and spreading sequences are not orthogonal, the \textit{near-far-problem} exists~\cite{TseBook}.
To solve this issue, we formulate pilot detection and channel estimation as a compressed sensing problem, where the sparse vector of length $J$ has only $K$ non-zero values.
Note that these values correspond to the channels between UEs and AP.
The sensing matrix is $\Pm$, and the goal is to recover a sparse vector with $R_m < K$ non-zero values.   
As a solver, we adopt the Orthogonal Matching Pursuit (OMP) algorithm~\cite{OMPanalysis,OMPwave}.
The idea behind of OMP is to peel the detected pilots from the received signal $\Ym_m^p$.
This approach has a flavor of Non-orthogonal Multiple Access (NOMA) decoding \cite{NOMA}.
In contrast to traditional OMP, the version of OMP employed in our receiver terminates the iterative process once it recovers $R_m$ indices. Consequently, the remaining $K-R_m$ indices are considered noise during the OMP iterations.
Recall that $\Ym_m^p$ is the pilot signal at the $m$th AP.
Let $i$ denote the iteration index of OMP.
Then, $\mathcal{S}_m^{(i)}$ is the set of the recovered indices at the $m$th AP at time $i$, and $S_m^{(i)} = |\mathcal{S}_m^{(i)}|$.
Note that $S_m{(i)} = i, \ \forall i = 1,2,\dots,R_m$. 
Define $\Ym_{\mathrm{resid}}^{(i)}$ as the residual obtained by subtracting the pilot sequences at the $i$th iteration from the pilot signal $\Ym_m^p$.
The first step of the algorithm is to compute the energy of each pilot sequence as
\begin{equation} \label{eq:EnergyDetector}
    \lambda_j = \Vert\Pm^*[:,j]\Ym_{\mathrm{resid}}^{(i)}\Vert^2_\mathrm{2}, \ \ \forall j \in[J],
\end{equation}
Then, the algorithm outputs the index with the largest energy
\begin{align*}
    \hat{\jmath} = \mathrm{argmax}_{j \in [J]}\lambda_j
\end{align*}
and updates the recovering set as $\mathcal{S}_m^{(i+1)} =  \mathcal{S}_m^{(i)} \cup \{\hat{\jmath}\}$.
The pilot signal $\Ym_m^p$ and the active pilots $\Pm[:,\mathcal{S}_m^{(i)}]$ are subsequently used to estimate the channel by solving the least squares problem,
\begin{align*}
  \hat{\Gm}_m^{(i)} =  \argmin_{\Gm_m \in \mathbb{C}^{S_m^{(i)} \times N}} \ \ \| \Ym_m^p - \Pm[:,\mathcal{S}_m^{(i)}]\boldsymbol\Pi_{(i)}^{\frac{1}{2}} \Gm_m \|_2,
\end{align*}
where $\boldsymbol\Pi_{(i)}^{\frac{1}{2}}$ is an $i \times i$ diagonal matrix with elements $P$.
It is straightforward to show that $\hat{\Gm}_m$ can be computed as
\begin{align*}
    \hat{\Gm}_m &= \Big(\Pm^*[:, \mathcal{S}_m]\boldsymbol\Pi_{(i)}\Pm[:,\mathcal{S}_m ]\Big)^{-1}\boldsymbol\Pi_{(i)}^{\frac{1}{2}}\Pm^*[:,\mathcal{S}_m] \Ym^p.
\end{align*}
The final step of the OMP iteration is to subtract the interference of the recovered pilots/estimated channels, and pass the residual to the energy detector for the next round \eqref{eq:EnergyDetector},
\begin{align*}
    \Ym_\mathrm{resid}^{(i+1)} = \Ym^p_m - \Pm[:, \mathcal{S}_m^{(i)}]\boldsymbol\Pi_{(i)} ^{\frac{1}{2}}\hat{\Gm}_m^{(i)}. 
\end{align*}
We continue this process until $R_m$ pilots are recovered.
\subsubsection{Symbol Estimation}
The next step of the receiver is the estimation of the symbols.
Given the information that is available to an AP, we can write the received data signal as
\begin{equation*}
\begin{split}
    \Ym^d_m &= \Dm[:,\mathcal{S}_m] \boldsymbol{\Pi}_{(R_m)}^\frac{1}{2} \hat{\Gm}_m \\
    &\quad + \Dm [:,\mathcal{S}_m]\boldsymbol\Pi_{(R_m)}^\frac{1}{2} \Big( \Gm_m - \hat{\Gm}_m  \Big)
     + \mathbf{I}^d_m + \Zm^d_m,
\end{split}
\end{equation*}
where $\Dm[:,\mathcal{S}_m]$ are the columns of $\Dm_a$ corresponding to the recovered users, and $\mathbf{I}^d_m$ is the interference of the remaining $K-R_m$ users.
We assume that the inference and the channel estimation error is small compared to thermal noise, and we perform LMMSE filtering to estimate the symbols.
Let us define the vectorized received signal as
\begin{align*}
    \underbrace{\begin{bmatrix}
        \Ym^d_m[\nv_t, 1]\\
        \Ym^d_m[\nv_t,2]\\
        \vdots\\
        \Ym^d_m[\nv_t,N]
    \end{bmatrix}}_{\tilde{\yv}_t \in \mathbb{C}^{L N \times 1}}
    = 
    \underbrace{\begin{bmatrix}
        \Am[:, \mathcal{S}_m] \boldsymbol\Lambda_{1,m}\\
        \Am[:, \mathcal{S}_m]\boldsymbol\Lambda_{2,m}\\
        \vdots\\
        \Am[:, \mathcal{S}_m]\boldsymbol\Lambda_{N,m}
    \end{bmatrix}}_{ \hat{{\Bm}} \in \mathbb{C}^{L N \times R}} \; \underbrace{\rv_t}_{R \times 1} +
    \underbrace{\begin{bmatrix}
        \Zm^d_m[\nv_t,1]\\
        \Zm^d_m[\nv_t,2]\\
        \vdots\\
        \Zm^d_m[\nv_t,N]
    \end{bmatrix}}_{\tilde{\zv}_t \in \mathbb{C}^{L N \times 1}}
\end{align*}
where $ \nv_t := [(t-1)L:tL]$,  $\boldsymbol\Lambda_{n,m} := \boldsymbol\Pi_{(R_m)}^\frac{1}{2}\mathrm{diag}(\hat{\Gm}_n)$, and $\rv_t$ is the vector contains the symbols of the recovered users at time $t$.
Then, the vectorized received signal becomes
\begin{equation*} 
    \tilde{\yv}_t =  \hat{\Bm} \rv_t + \tilde{\zv}_t .
\end{equation*}
The LMMSE estimates of the $R_m$ users at time $t \in [T]$ are given by
\begin{equation*}
\hat{\rv}_t = \hat{\Bm}^*\left(  \hat{\Bm}\hat{\Bm}^* +\sigma^2{\mathbf I}_{LN} \right)^{-1} \tilde{\yv}_t, \ \forall \quad t \in [T],
\end{equation*}
where ${\mathbf I}_{LN}$ is the $LN \times LN$ identity matrix.
Let $\{ \hat{\rv}_{t,m}\}_{t=1}^T$ be set that contains the estimates of the symbols detected at the $m$th AP. 
The information transmitted from the $m$th AP to the CPU is denoted by $\mathtt{p}_m = \mathcal{S}_m \cup \{ \hat{\rv}_{t,m}\}_{t=1}^T$.
We note that, in the Cell-Free \textit{Level~2} cooperation method, only the symbol estimates are sent to the CPU.
However, since the index of the active pilots carries important information, it is necessary to transmit $\mathcal{S}_m$.

\subsection{Central Processing Unit (CPU)}
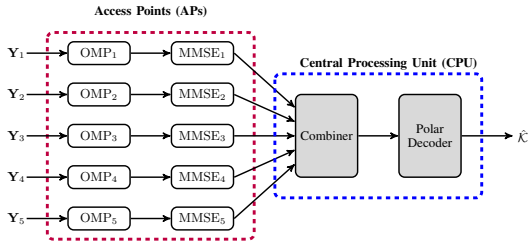
\begin{figure}[h]
    \centering
    \scalebox{0.55}{\begin{tikzpicture}
  [
  font=\small, draw=black, >=stealth', line width=1pt,
  channel/.style={rectangle, minimum height=20mm, minimum width=15mm, draw=black, fill=gray!30, rounded corners},
  encoder/.style={rectangle, minimum height=5.5mm, minimum width=15mm, draw=black, fill=gray!10, rounded corners},
  decoder/.style={rectangle, minimum height=20mm, minimum width=15mm, draw=black, rounded corners},
  message/.style={rectangle, minimum height=5.5mm, minimum width=15mm, draw=black, rounded corners}
  ]

\foreach \e in {1,2,3,4,5} {
  \node[message] (e\e) at (2.5,0.0-\e) {{$\mathrm{MMSE}_\e$}};
}

\draw[draw=purple, dashed, line width=2.0pt,rounded corners] (-1.25,-0.5) rectangle (3.75,-5.5);

\foreach \m in {1,2,3,4,5} {
  \node[message] (m\m) at (0.0,0.0-\m) {${\mathrm{OMP}_\m}$}
  edge[->] (e\m);
  \draw[<-] (m\m) -- (-1.75,0.0-\m);
  \node at (-2,0.0-\m) {$\mathbf{Y}_\m$};
}

\node[channel,align=center] (channel) at (5.5,-3) {Combiner};
\node[channel,align=center] (decoder) at (8,-3) {Polar\\Decoder};
\draw[->] (channel) -- (decoder);

\draw[->] (decoder.east) -- (10,-3);
\node at (10.25,-3) {$\hat{\mathcal{K}}$};

\draw[->] (e1.east) -- (channel);
\draw[->] (e2.east) -- (channel);
\draw[->] (e3.east) -- (channel);
\draw[->] (e4.east) -- (channel);
\draw[->] (e5.east) -- (channel);

\node at (1.25,0) {\textbf{Access Points (APs)}};

\node at (7,-1.25) {\textbf{Central Processing Unit (CPU)}};

\draw[draw=blue, dashed, line width=2.0pt,rounded corners] (4.25,-1.5) rectangle (9.25,-4.5);

\end{tikzpicture}}
     \caption{Example of $5$ APs and a CPU. The block diagram shows the main blocks of the two processing units.}
    \label{fig:APsCPU}
\end{figure}

The CPU is the last processing step in a CF system.
The role of this block is to combine the estimates from different APs, and to run a list-polar decoder to recover the second part of the message.
Let $\hat{\Sm}_m$ be a ${T \times R_m}$ matrix with rows given by $\hat{\rv}_{1,m}\T, \hat{\rv}_{2,m}\T,\dots, \hat{\rv}_{T,m}\T$.
The symbols estimates of the UE with the sequence $j$ is given by
\begin{align}
    \hat{\sv}_j =  \sum_{m=1}^M\mathbf{1}_{\{j \in \mathcal{S}_m\}} \hat{\Sv}_m[:, f_m(j)]
\end{align}
where  $\mathbf{1}_{\{j \in \mathcal{S}_m\}}$ is the indicator function and
$f_m(j)$ is a function that maps the index of the sequence to the column of $\hat{\Sm}_m$.
Here, we assume that the probability of collision is small and we can associate a user with an index.
The estimates $\hat{\sv}_j$, are passed to a list-polar decoder. 
A block diagram of APs and CPU is illustrated in Fig.~\ref{fig:APsCPU}.

\section{Simulation Results}

To demonstrate the performance of the proposed scheme\footnote{The source code for the CEFURA communication scheme is available at \url{https://github.com/EngProjects/mMTC}.}, we compare a traditional network with one AP in the middle of the cell and a CF setup wherein APs are placed in a square grid.
We show the behavior of the scheme for different system parameters.
Specifically, we highlight how the cell size affects the performance of the centralized and CF systems.
We also explore how the distribution of the UEs influences the error rate of these systems.
The parameters used for simulations appear in Table~\ref{table:par}.
The total number of channel uses is $n=3200$, and each AP aims to recover $R_m = 7, \forall m \in [M]$ UEs.
We assume that the product of the number of APs $M$ and the number of recovered users $R_m$ is greater than $K$.
This is equivalent to having every user connected to at least one AP in a traditional CF setting.
\begin{table}[h]
\begin{center}
\caption{This is a summary of the simulation parameters.}
\begin{tabular}{ | c| c| c | c| c | c| c| c| c|} 
  \hline
   $R_m $ & $\sigma^2$ & $P_k $ & $B$ & $B_f$ & $B_\mathrm{crc}$ & $n_c$ & $n_p$ & $L$ \tabularnewline
  \hline
   7 & -84~dBm &  10~mW & $100$ & $15$ & $16$ & $512$ & $640$ &$10$ \tabularnewline 
  \hline
\end{tabular}
  \end{center}

\label{table:par}
\end{table}
However, if the number of users is greater than $MR_m$, then additional APs can be added to satisfy $K \leq MR_m$.
We stress that the system is scalable because $R_m$ does not scale with $K$.
The large scale coefficients are generated from the model in \eqref{eq:LSC}.

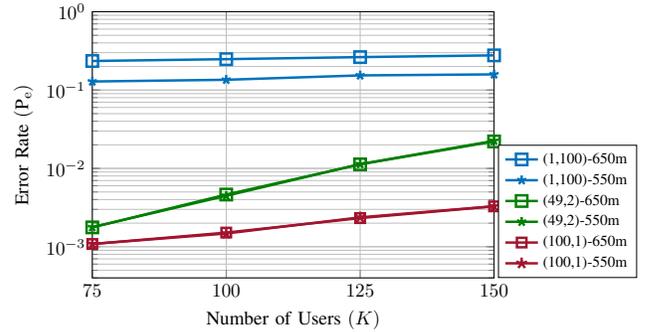
\begin{figure}[h]
\centering
\scalebox{0.8}{\begin{tikzpicture}[scale=0.9]
\definecolor{mycolor1}{rgb}{0.63529,0.07843,0.18431}%
\definecolor{mycolor2}{rgb}{0.00000,0.44706,0.74118}%
\definecolor{mycolor3}{rgb}{0.00000,0.49804,0.00000}%
\definecolor{mycolor4}{rgb}{0.87059,0.49020,0.00000}%
\definecolor{mycolor5}{rgb}{0.00000,0.44700,0.74100}%
\definecolor{mycolor6}{rgb}{0.74902,0.00000,0.74902}%

\begin{axis}[
ymode=log,
xmode=normal,
ymin=4e-4,
ymax=1,
xmin=75,
xmax=150,
xtick = {50,75,...,150},
grid=both,
width=9cm,
height=6.5cm,
ylabel={Error Rate $(\pe)$},
xlabel={Number of Users $(K)$},
legend style={at={(1.01, 0)},anchor=south west,draw=black, fill=white, legend cell align=left,font=\footnotesize}]

\addplot [color=mycolor2,solid,line width=1.3pt,mark size=3pt,mark=square,mark options={solid}]
  table[row sep=crcr]{
50 0.228733333333333 \\
75  0.234987978142076\\
100 0.248058797844667 \\
125 0.263522657539384\\
150 0.277456224524991\\
};
\addlegendentry{(1,100)-650m};

\addplot [color=mycolor2,solid,line width=1.3pt,mark size=2.5pt,mark=star,mark options={solid}]
  table[row sep=crcr]{
50 0.117733333333333\\
75 0.128671973466003\\
100 0.135333333333333\\
125 0.153935913100081\\
150 0.158584850045205\\
};
\addlegendentry{(1,100)-550m};

\addplot [color=mycolor3,solid,line width=1.3pt,mark size=3pt,mark=square,mark options={solid}]
  table[row sep=crcr]{
50 0.0011\\
75 0.00176649122807016 \\
100 0.004635\\
125 0.0112639999999998\\
150  0.0220833333333334\\
};
\addlegendentry{(49,2)-650m};

\addplot [color=mycolor3,solid,line width=1.3pt,mark size=2.5pt,mark=star,mark options={solid}]
  table[row sep=crcr]{
50 0.000889411764705883\\
75 0.00177333333333332\\
100 0.004465\\
125 0.0113359999999998\\
150 0.0224966666666668 \\
};
\addlegendentry{(49,2)-550m};

\addplot [color=mycolor1,solid,line width=1.3pt,mark size=2.5pt,mark=square,mark options={solid}]
  table[row sep=crcr]{
50 0.001198954248366\\
75 0.00108409356725145\\
100  0.00151490099009899\\
125 0.00231996825396825\\
150 0.00327666666666671\\
};
\addlegendentry{(100,1)-650m};

\addplot [color=mycolor1,solid,line width=1.3pt,mark size=3pt,mark=star,mark options={solid}]
  table[row sep=crcr]{
50 0.00098954248366013\\
75 0.00108409356725145\\
100  0.00147499999999998\\
125 0.002372\\
150 0.00329000000000004\\
};
\addlegendentry{(100,1)-550m};

\end{axis}

\end{tikzpicture}%
}
 \caption{Performance of the scheme for different cell size and $(M,N)-D$ values.}
\label{fg:D2C}
\end{figure}

\noindent
\textbf{Different Cell Size:} 
Figure~\ref{fg:D2C} compares the performance of CEFURA to that of a centralized system.
We consider $K \in \{75,100,125,150\}$ UEs where the location of the users follows a binomial Point Process (PP) on a $[0,C] \times [0,C]$
microcell~\cite{3GPP-R16}, with $C \in \{550,650\}$.
Results indicate that distributing antennas improves overall performance.
The CF system with 100 APs and $N=1$ start to outperform its CF counterpart with 49 APs, each with two antennas.
To understand this phenomenon, let us focus on a particular AP.
As $K$ increases, the density in the vicinity of this AP increases, on average.
As a result, interference at this AP raises and thus the performance decreases.
In the $100$ APs case, the Voronoi region of each AP is smaller and, hence, interference is limited.

\begin{figure}
\centering
\scalebox{0.8}{\begin{tikzpicture}[scale=0.9]
\definecolor{mycolor1}{rgb}{0.63529,0.07843,0.18431}%
\definecolor{mycolor2}{rgb}{0.00000,0.44706,0.74118}%
\definecolor{mycolor3}{rgb}{0.00000,0.49804,0.00000}%
\definecolor{mycolor4}{rgb}{0.87059,0.49020,0.00000}%
\definecolor{mycolor5}{rgb}{0.00000,0.44700,0.74100}%
\definecolor{mycolor6}{rgb}{0.74902,0.00000,0.74902}%

\begin{axis}[
ymode=log,
xmode=normal,
ymin=4e-4,
ymax=0.1,
xmin=75,
xmax=150,
xtick = {75,100,...,150},
grid=both,
width=9cm,
height=6.5cm,
ylabel={Error Rate $(\pe)$},
xlabel={Density $(\mu)$},
legend style={at={(1.01, 0)},anchor=south west,draw=black, fill=white, legend cell align=left,font=\footnotesize}]

\addplot [color=mycolor3,solid,line width=1.3pt,mark size=2.4pt,mark=square,mark options={solid}]
  table[row sep=crcr]{
75 0.00693710582784575\\
100 0.0177 \\
125 0.0406553716580165\\
150 0.0674313556477207\\
};
\addlegendentry{(49,2)-Thomas};

\addplot [color=mycolor3,solid,line width=1.3pt,mark size=2.4pt,mark=o,mark options={solid}]
  table[row sep=crcr]{
75 0.00509000513684721\\
100 0.0159472320482354\\
125 0.0343556495469184\\
150 0.0579968655065554\\
};
\addlegendentry{(49,2)-Matérn};

\addplot [color=mycolor3,solid,line width=1.3pt,mark size=2.4pt,mark=star,mark options={solid}]
  table[row sep=crcr]{
75 0.00199480924023753\\  
100 0.00485902550513249\\ 
125 0.011388000221557\\ 
150  0.0231407395626478\\ 
};
\addlegendentry{(49,2)-Poisson};

\addplot [color=mycolor1,solid,line width=1.3pt,mark size=2.4pt,mark=square,mark options={solid}]
  table[row sep=crcr]{
75 0.00159291329331272 \\ 
100 0.00342774696535437 \\
125 0.00718470380116968\\
150 0.0137816102209361\\
};
\addlegendentry{(100,1)-Thomas};

\addplot [color=mycolor1,solid,line width=1.3pt,mark size=2.4pt,mark=o,mark options={solid}]
  table[row sep=crcr]{
75 0.00170793813813706\\
100 0.00285389616045898\\
125 0.00551258702182526\\
150 0.0102570237954681\\
};
\addlegendentry{(100,1)-Matérn};

\addplot [color=mycolor1,solid,line width=1.3pt,mark size=2.4pt,mark=star,mark options={solid}]
  table[row sep=crcr]{
75 0.00124147922603374\\ 
100 0.00163414487782313\\
125 0.00228288233908598\\  
150 0.00312999802399038\\ 
};
\addlegendentry{(100,1)-Poisson};
\end{axis}

\end{tikzpicture}%
}
 \caption{Comparison of various Point Processes for modeling the spatial distribution of UE under different $(M, N)-$Distribution}
\label{fg:clusterSim}
\end{figure}
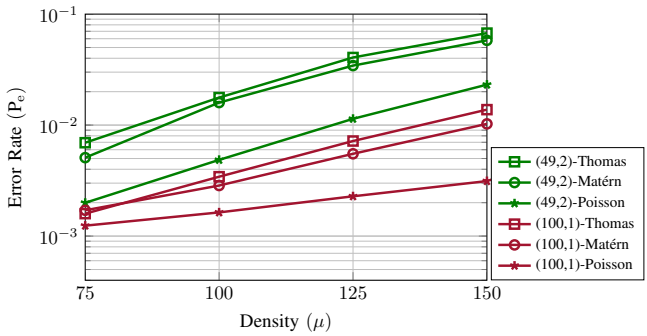

\noindent
\textbf{Different UE Distributions:}
Consider a $650 \times 650 \ \mathrm{m}^2$ geographic area, with two CF configurations, and different UE distributions.
The locations of the UEs follow a \textit{Poisson}, \textit{Thomas}, or \textit{Mat\'{e}rn} PP.
The number of UEs is random, yet we assume local conditions are available at each decoder.
Also, let $c$, $\lambda D^2$, and $\tau$ be the number of clusters, the mean of the parent process, and that of the daughter process (if applicable), respectively.
Let $c = \lambda D^2$ and $\tau = \frac{K}{c}$.
For PPP, we set $\tau=1$ and $c=K$.
Figure~\ref{fg:clusterSim} illustrates the performance of the proposed scheme for $c=25$.
We note that the average density $\mu$ is equal to $K$.
As expected, when the UEs follow either a \textit{Thomas} or \textit{Mat\'{e}rn} PP, the performance suffers slightly.
Nevertheless, the trends in terms of AP density remain.


\section{Conclusion}
This article considers the massive MIMO unsourced random
access problem on a quasi-static Rayleigh fading channel, in a cell-free architecture.
We propose a communication scheme that can operate in such a scenario.
Numerical results show that by combining cell-free and URA the performance of the system can be improved compared to centralized URA.
The main difference between the Massive MIMO quasi-static fading URA channel and the cell-free architecture is; 1) The coefficient of the channel between an UE and an AP is not independent and the structure of the antenna array has to be defined in order to compute the correlation between them.
2) 

\bibliographystyle{IEEEbib}
\bibliography{referencesCellFree, referencesFASURA }

\end{document}